\documentclass[nobm]{epl}

\title{Small-world behavior in a system of mobile elements}
\shorttitle{Dynamic Small-Worlds}
\author{Susanna C. Manrubia\inst{1} \and Jordi Delgado\inst{2} 
\and Bartolo Luque\inst{3}}
\shortauthor{S.C. Manrubia \etal}
\institute{
  \inst{1} Max Planck Institute for Colloids and Interfaces, Theory 
Division.\\ 14424 Potsdam, Germany \\
  \inst{2} Departament de Llenguatges i Sistemes
Inform\`atics, Universitat Polit\`ecnica de \\ Catalunya. Campus Nord,
M\`odul C6, 08034 Barcelona, Spain \\
  \inst{3} Centro de Astrobiolog\'{\i}a (CAB), INTA. Carretera de Ajalvir 
Km. 4, 28850. \\ Torrej\'on de Ardoz, Madrid, Spain}


\begin{document}
%\verb|nobm|
\maketitle

\begin{abstract}
We analyze the propagation of activity in a system of mobile automata. 
A number $\rho L^d$ of elements move
as random walkers on a lattice of dimension $d$, while with a small
probability $p$ they can jump to any empty site in the system. We show that
this system behaves as a Dynamic Small-World (DSW) and present analytic 
and numerical results for several quantities. Our analysis shows that the 
persistence time $T^*$ (equivalent to the persistence size $L^*$ of 
small-world networks) scales as $T^* \sim (\rho p)^{-\tau}$, with $\tau = 
1/(d+1)$.
\end{abstract}

The network of collaboration of actors in movies, of world airports, or of 
neural connections in the worm {\it Caenorhabditis 
elegans}, can be represented in the 
form of a graph.  Persons, airports, and neurons stand for the 
nodes of the network, and a bond is drawn when two actors are cast 
together in a film, when a non-stop flight exists between two airports, or 
when a neural connection is present, respectively \cite{Watts,Classes}. 
The relational graphs resulting from these apparently disparate systems have 
two interesting properties that qualify them as {\it small-world} graphs: 
They are highly clustered at the local scale (similar to what happens in 
regular graphs), and the average number of bonds $\bar \ell$ that separate 
two randomly chosen nodes in a graph of size $L$ scales as $\bar{\ell} \sim 
\log L$ (a characteristic property of random graphs).

Small-world (SW) graphs efficiently interpolate between regular and random 
graphs thanks to a small number $p L^d$ of {\it shortcuts} (long-range
connections, $p \ll 1$) which are 
superimposed on a regular lattice formed by $L^d$ sites. Much attention has
been devoted recently to the topological properties of such graphs, which 
determine the characteristic path length $\bar \ell$ and its scaling 
properties \cite{Top,pers}, and to the effect that a small-world-like 
connectivity might have on the properties of dynamical 
systems~\cite{Dyn,Dyn2}.
Up to now, the attention has been only focused on this relevant type of 
networks, SW. Nonetheless, the restriction of 
activity propagating {\it on a quenched, undirected graph}, where elements 
are fixed in their positions and information propagates in a symmetric way 
(from element $A$ to element $B$ as well as {\it vice versa}), is a strong 
one, and perhaps not always the best representation of a world formed by 
mobile elements where connections (relational bonds) appear and disappear 
in the course of time. In this sense, recent approaches have tried to model 
the formation and evolution of the topology in SW and other networks throught 
slow
changes in their bonds \cite{Bara,Doro}. Our focus here is not in this large 
time-scale, but on the shorter time-scale in which the movement of the 
elements in the physical space is essential in defining contacts between 
nodes. As an example, consider the case of propagation of a 
tropical disease. Such events usually start in a ``far-away'' place (in the 
Euclidean sense), from which the carrier of the disease flies away. 
He or she arrives at a new destination and transmits the disease, and the 
process repeats with the new carriers. This is a clear example of a directed,
dynamical graph, with moving elements which create and remove connections
in a partly stochastic fashion. Indeed, in all cases where the carrier is not 
taking active part in the transmission of the signal, random encounters might 
play a main role when compared to the fixed network of acquaintances.

With the aim of analyzing the effect of movement in the transmission of a 
signal, we introduce a simple model for the formation of a Dynamic
Small-World (DSW), and study the propagation of activity among mobile elements
in the system. We define quantities analogous to the characteristic path 
length $\bar \ell$ and to the clustering coefficient $C$ in order to quantify 
the similarities with SW. Our analytical and numerical calculations show that 
DSW are quantitatively different from (quenched) SW.

The model is defined as follows. Consider a lattice formed by $L^d$ sites 
embedded in a $d-$dimensional space, where $\rho L^d$ automata can move.
The density of automata $\rho$ is a parameter of the model. A time step 
consists of one move of all the active elements in the system plus the 
synchronous transmission of activity to new inactive neighbours. The precise 
implementation of these rules is as follows: i) Choose one of the active 
elements according to a random sequential updating (i.e. each active element
will be updated once and only once in an uncorrelated fashion with respect
to the previous move). With probability $p$, the automaton 
selects an empty site in the system and jumps there. With the complementary 
probability $1-p$, it looks for a free site in its neighbourhood and moves to 
it. The neighbourhood consists of the $2d$ adjacent sites. ii) Once all the 
active automata have moved, activity is simultaneously transmitted by each
active element to any other inactive element in its neighbourhood.
As initial condition an automaton is selected at random and activated at 
$t=0$, the rest of the system remains inactive. The active element starts
moving according to rule i). When an inactive element is met at a 
neighbouring site, activity is transmitted according to ii). We use periodic 
boundary conditions in our simulations.

The probability $p$ of jumping to a far-away
site is the equivalent of the density of shortcuts in SW. In that
case, the activity would propagate to one new bond per time step, and one could
define a velocity of propagation $v_{\rm SW}=1$\cite{Top}. In the DSW case, 
the time required to transmit the activity to the chosen element 
depends on the density of automata $\rho$, so we expect $v_{\rm DSW} = 
v(\rho) \le 1$.

\begin{figure}[tbp]
\twofigures[width=6.6cm,height=6cm]{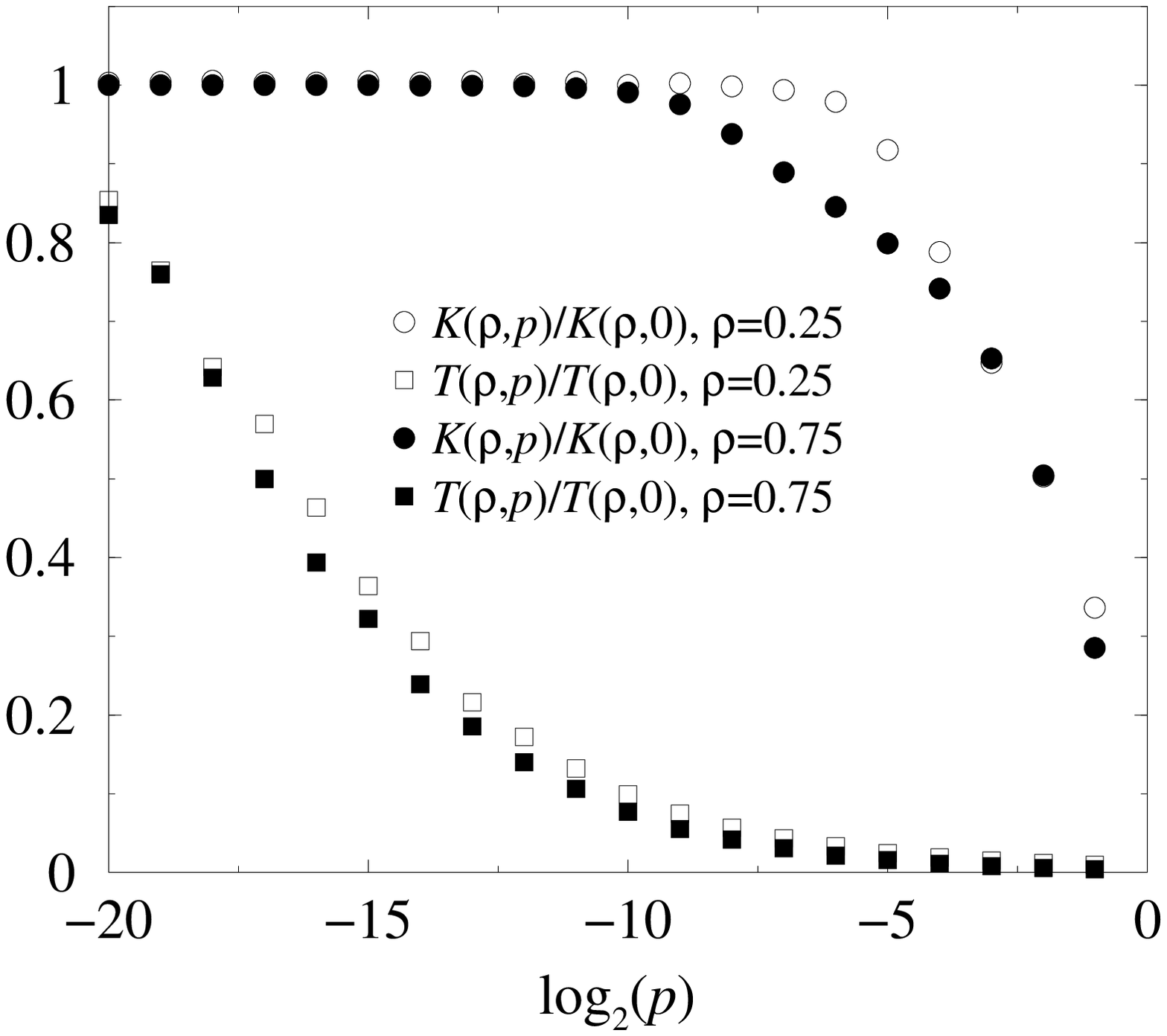}{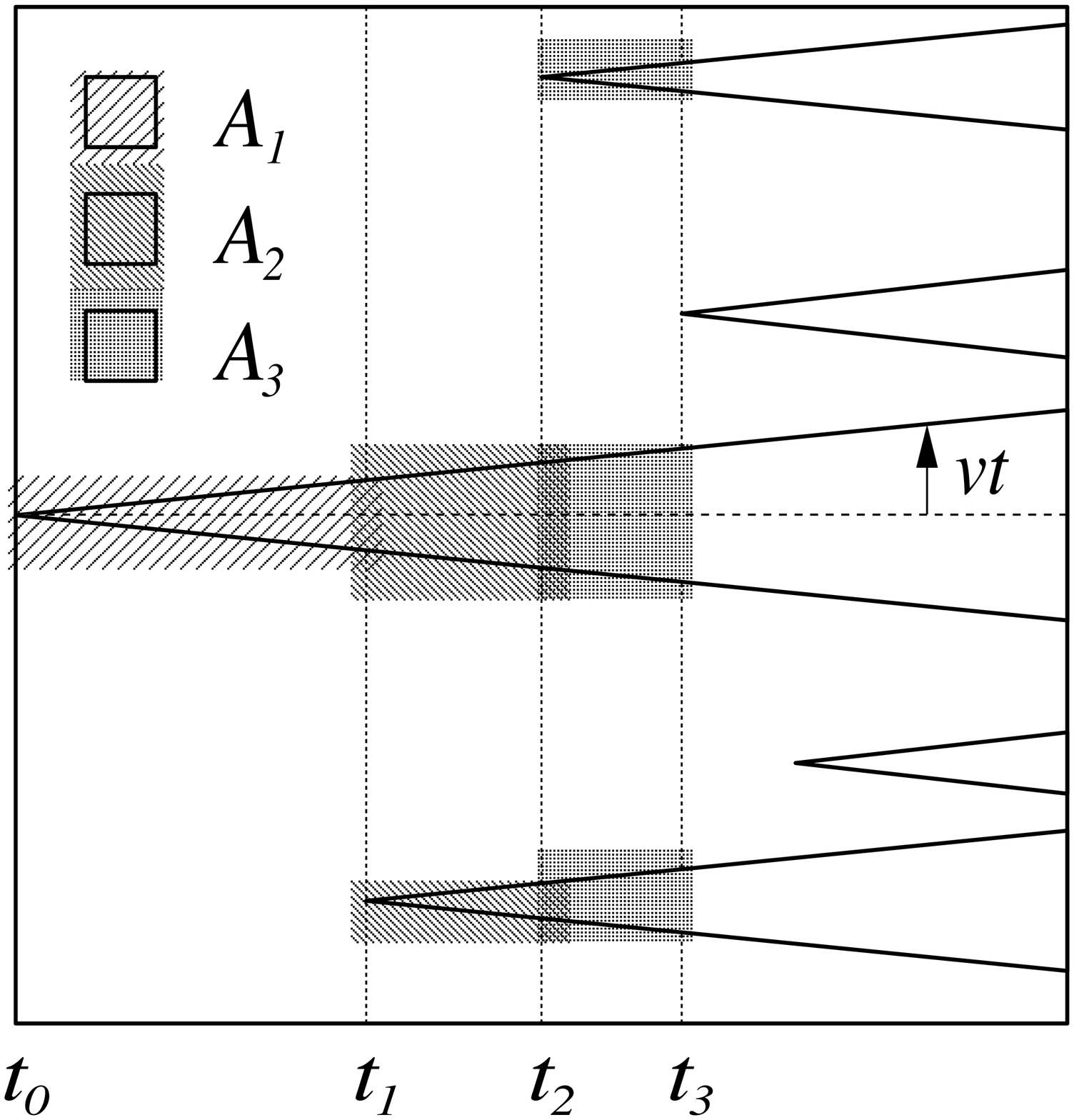}
\caption{Normalized characteristic time $T(\rho, p) / T(\rho, 0)$ and 
normalized neighbouring coefficient $K(\rho, p) / K(\rho, 0)$ as a function 
of $p$ for two different values of the density, as shown in the legend, in a 
one-dimensional system of size $L=5000$ (averages over $10^3$ independent 
realizations have been performed).}
\label{fig1}
\caption{Schematic representation of the propagation of activity in a Dynamic
Small-World. The initial condition corresponds to a single active particle
which moves as a random walk. The spatio-temporal triangle signaling the 
spreading of activity has a half-angle $\alpha = \arctan v$. A second 
triangle starts the first time that an automaton jumps out of $A_1$, at a time 
$t_1=(v \rho p)^{-1/2}$. When $A_k=A_1$, the $k$th jump takes
place.}
\label{fig2}
\end{figure}

We start by defining two quantities analogous to the characteristic path length
$\bar \ell (p)$ and the clustering coefficient $C(p)$ of SW 
(for details, see Ref. \cite{Watts}).
In our case, the quantity equivalent to $\bar \ell$ 
is the characteristic time $T (\rho, p)$ for the activity starting at 
the seed to reach a randomly chosen element in the system, averaged
over many independent realizations of the
process. Because $v_{\rm SW}=1$ in SW, $\bar \ell$ and $T$ coincide.
We will call the analogous of the clustering coefficient {\it neighbouring 
coefficient}, and define it in the following way. Consider the set 
of particles located in the neighbourhood of element $i$ at time $t$, 
${\cal{N}}_t(i)$. At the next time step, this set will be in principle 
different, since all active elements have tried to move to new sites. We 
define the neighbouring coefficient for element $i$ at time $t$ as $K_t(i; 
\rho, p)= (2d)^{-1} {\cal{U}}_t(i)$, where 
${\cal{U}}_t(i) \equiv \textrm{card} \{{\cal{N}}_t(i) \cap
{\cal{N}}_{t-1}(i) \}$. The average neighbouring 
coefficient for a system with parameters $p$ and $\rho$ 
is $K (\rho, p) = \langle (\rho L^d)^{-1} \sum_i  K_t(i; \rho, p) \rangle$,
where the brackets $\langle . \rangle$ represent a temporal average. 
Our numerical simulations in $d=1, \; 2$ reveal 
that the small-world behavior of our system is well described through these
two quantities (see Fig. \ref{fig1}). Indeed, for a wide range of $p$ values, 
the local configuration is maintained (the value of $K$ is close to unity), 
while the typical number of time steps required to reach a randomly chosen 
automaton from the initial source is small and scales as $\log L$. This 
behavior is characteristic of SW systems.

Our system admits a geometrical description suitable for calculating 
several quantities analytically. We will focus our interest on the 
growth of the volume $V(t)$ occupied by $a(t)$ active elements 
as a function of time $t$. There is a simple
relation between the active volume and the number of active particles:
$\rho V(t) = a(t)$. In the following, we restrict our calculations
to the case $d=1$. Some generalizations to arbitrary $d$ are
presented at the end of the letter. 

At $t_0=0$, only one element is active. 
This element moves as a random walker during about $t \sim \rho^{-2}$ time 
steps (the average distance between elements is of order $1/\rho$),
until it meets a second element at a neighbouring site, and 
activity is transmitted. This kind of spreading of activity persists for 
$t_1$ time steps, 
where $t_1$ signals the jump of the first active element
to a domain disconnected (in terms of activity) from the previous one. 
For $p^{-1/2} \gg \rho^{-2}$ (contact through
random walker moves is much more frequent than contact through long-distance
jumps, see below), we can use a continuous approximation to the problem, 
as shown schematically
in Fig. \ref{fig2}. The vertical axis represents physical distance in a
one-dimensional space, and the horizontal axis represents time. At time 
$t$, active elements cover a volume $V(t)$ equal to the sum of the lengths
obtained from the intersection of a
vertical line drawn at $t$ with the triangles.  For the particular case 
$p=0$ the activity propagates at a constant velocity $v(\rho)$ and, in
the geometrical representation, only one triangle is present.

We can easily calculate the average time $t_1$
when the first jump (or shortcut) takes place. To this end, we make the
annealed approximation $t \propto \langle \sqrt{n(t)} \rangle
\sim \sqrt{\langle n(t) \rangle}$ (where $n(t)$ is the total number
of attempted jumps up to time $t$), and obtain

\begin{equation}\label{t1}
t_1 = \sqrt{1 \over v \rho p} \; .
\end{equation} 
According to the picture in Fig. \ref{fig2} we can also calculate the times 
for the appearance of the next jumps,

\begin{equation}\label{t2}
t_2 = {1 + \sqrt 3 \over 2} t_1 \;\;\;\;\;\;\;\;\;\;\;\;\;
t_3 = {1 \over 2} \left[  1 + {\sqrt 3 \over 3} + {2 \sqrt{6} \over 3}
 \right] t_1 \; ,
\end{equation}
and so on.
Note that jumps are not equidistant in time, and that
they happen more and more frequently for longer times: $(t_1-t_0) > (t_2 - t_1)
> \dots > (t_{k+1}-t_k) > \dots$, and these differences go to zero for $k \to
\infty$. 
In general, we can write an expression for the time by which $k$ jumps 
have taken place,

\begin{equation}\label{kk}
t_1^2 k = \sum_{i=0}^{k-1} (t_k - t_i)^2 \; .
\end{equation}
If we consider the new variable $\tau_k = (t_k - t_{k-1})/t_1$ the difference
between consecutive times satisfies

\begin{equation}\label{ittau}
k \tau_k^2 + 2 \tau_k \; \sum_{i=1}^{k-1} i \tau_i =1 \; .
\end{equation}
We look for a solution of the form
%\begin{equation}
$\tau_k = (a + \epsilon_{k-1} - \epsilon_k)/k$.
%\end{equation}
Direct substitution into Eq. (\ref{ittau}) yields $a = {1 / \sqrt{2}}, \;
\epsilon_0 = {a/2}$, and $\epsilon_k \sim {b'/ k}$, where $b'$ is 
obtained by imposing a final condition for $k \to \infty$. 
In the original discrete variables, we obtain now

\begin{equation}\label{tk}
t_k =  t_1 \sum_{i=1}^k \tau_i \simeq t_1 \left( {1 \over \sqrt{2}} 
\sum_{i=1}^k {1 \over i} + b - {c \over k^2} \right) \; .
\end{equation}
The coefficients of this expression are numerically obtained by fitting the
iterative solution of Eq. (\ref{ittau}) to the expression above. This returns
$b=0.3124 \dots , \; c= 0.0337 \dots$.

\begin{figure}[t]
\twofigures[width=6.6cm]{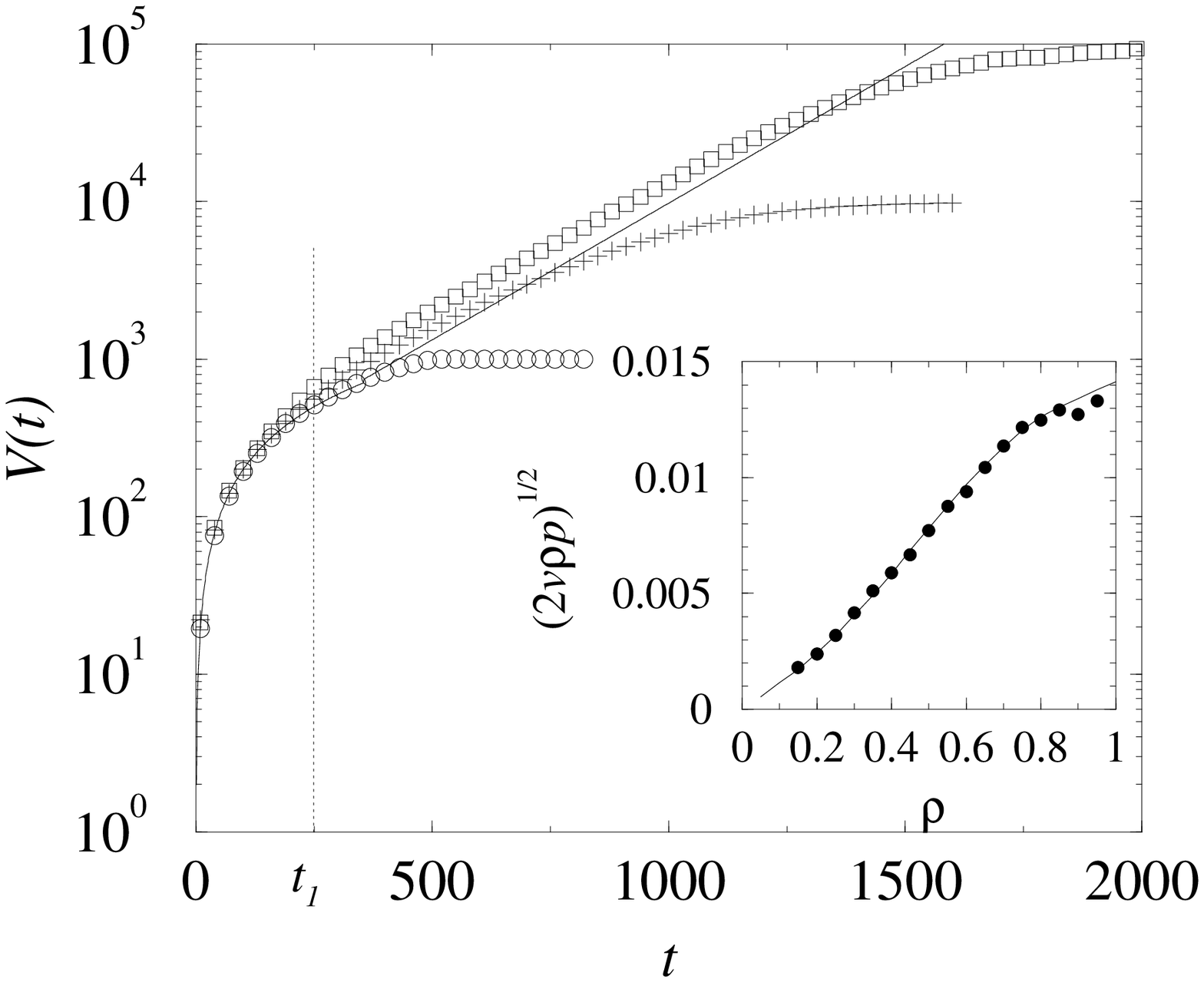}{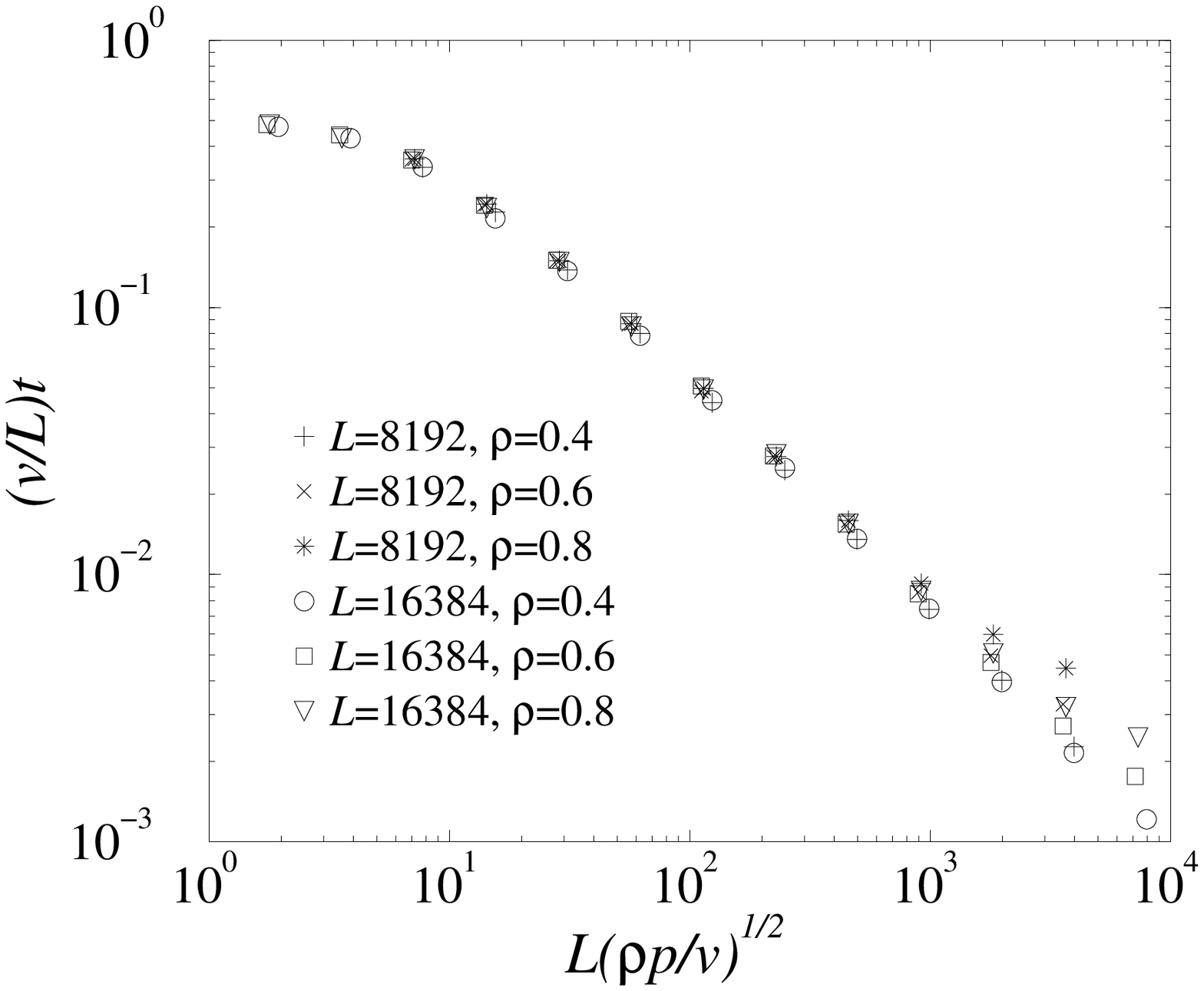}
\caption{Volume $V(t)$ covered by active particles.
Main plot: The continuous line 
corresponds to $2vt$ for $t<t_1$ and to our solution 
Eq. (\ref{Solvol}) for $t>t_1$. These simulations were made 
for systems of size $L=10^3$ (circles), $10^4$ (crosses), and $10^5$ (squares)
with $\rho = 0.8$ and 
$p=10^{-5}$. Inset: In the limit $t \gg t_1$ and $L \to \infty$, $V(t)$ grows 
exponentially at a rate $\sqrt{2 v \rho p}$. This result (continuous line,
with velocity numerically computed at $p=0$), is compared with numerical 
interpolations for systems of size $L=10^5$ (circles), with $\rho=0.4$ and
$p=10^{-4}$ in the exponential regime.}
\label{fig3}
\caption{Data collapse for the characteristic time $T$ as a function of 
the rescaled system parameters. The simulations were performed in dimension 
$d=1$ and averaged over 100 independent realizations. Other parameters
are as shown in the legend. There are corrections to scaling for $p 
\to 1$, where the assumption $p^{-1/2} \gg \rho^{-2}$ breaks down.}
\label{fig4}
\end{figure}

For time $t<t_1$, the volume grows linearly, $V(t<t_1) \simeq 2 v t$. We can 
also define the volume at the $k$th jump (for $t_k \ge t_1$) as

\begin{equation}\label{vol}
V(k) = 2 v \sum_{i=0}^{k-1} (t_k - t_i) = 2 v \left( k t_k
  -\sum_{i=1}^{k-1} t_i \right) \; .
\end{equation}
It follows from our previous calculations that

\begin{equation}
{V(k) \over 2 v t_1} = \sum_{i=1}^k i \tau_i = \sum_{i=1}^k
(a + \epsilon_{k-1} - \epsilon_k) = a k + \epsilon_0 - \epsilon_k \; .
\end{equation}
Inverting Eq. (\ref{tk}) and substituting above, we get the asymptotic 
expression

\begin{equation}\label{Solvol}
V(t) = \sqrt{2} v t_1 \exp \left\{ \sqrt{2} 
\left[ {t \over t_1} - \left( {{\bf C} \over \sqrt{2}} + b \right) \right] 
\right\} + O (t^{-1})\; ,
\end{equation}
where {\bf C}$=0.577215 \dots$ is Euler's constant.
Our numerical and analytical results are compared in Fig. \ref{fig3}. 

In these calculations, we have assumed that the time by which 
$k$ jumps have taken place is a deterministic variable defined by $t_k
\sim \sqrt{\langle n(t_k) \rangle}$, thus discarding fluctuations in $n(t)$.
If we calculate the fluctuations in the jumping process, it turns out that 
their
characteristic size is proportional to the approximate value $t_k$. The 
annealed approximation is quite good for $\rho$ close to unity and $p \to 0$, 
but the effect of the fluctuations is usually visible in the fact that 
the apparent velocity at $t<t_1$ is larger than $v(\rho)$ when $p > 0$
(this explains the small discrepancy seen in Fig. \ref{fig3} for the largest 
size). The bending at finite $V(t)$ observed in Fig. \ref{fig3} is a finite 
size effect: $V(t)$ saturates at the system size. The exponential regime is 
nonetheless well captured by our approach: In an infinite system, activity 
would propagate at a rate $\sqrt{2}/t_1 = \sqrt{2 v \rho p}$, as shown in the 
inset of Fig. \ref{fig3}. 

In the light of the above results, we have studied the scaling properties of 
the characteristic time $T(\rho, p)$ with the system size $L$, the 
jumping probability $p$, and the density of automata $\rho$, in a way 
similar to what has been done for SW \cite{NW}. First, we need to estimate 
the dependence of a persistence time $T^* \equiv t_1$ \cite{pers} on the 
parameters of the system. 
Despite having a whole spectrum of characteristic times, as given by 
(\ref{tk}), all of the $t_k$ scale in the same
form as $t_1$. Hence, we conjecture that $T^* \propto (v \rho p)^{-1/2}$ 
in $d=1$. Another relevant quantity with dimensions of time is $L/v(\rho)$. 
We propose the following scaling for $T$,

\begin{equation}
T \propto {L \over v(\rho)} f \left( L \left({\rho p \over v(\rho)}
\right)^{1/2} 
\right) \; ,
\label{scal}
\end{equation}
where $f(z)$ is a function of the dimensionless parameter $z=L/L^*$, and
we have defined the persistence size to be $L^* = v T^*$. We show in Fig. 
\ref{fig4} the resulting data collapse for several $T(\rho,p)-$ curves,
where $T(\rho,p)$ is the (minimum) time required for the activation of 
all particles in the system. Our numerical results support the scaling 
ansatz (\ref{scal}).

This result can be extended to dimensions $d>1$. Our previous 
calculations rely on the determination of a persistence time $T^*$,
which can be estimated in any dimension $d$ from a geometrical 
representation of the system. 
In general, we can talk of activity propagating in a $d-$dimensional lattice, 
at a velocity 
$v(\rho)$, from a source placed at the center of a hyper-sphere of radius 
$r(t) = v(\rho) t$. The first 
jump in this $d-$dimensional space will take place when the volume of the 
$(d+1)-$dimensional cone (where time has been included as an
additional dimension) contains $1/p$ automata. 
That is to say, the persistence time is defined by

\begin{equation}
1 = 
{2 \pi^{d/2} \over \Gamma(d/2)} v^d(\rho) \rho p \int_0^{T_d^*} {t^d} dt \; ,
\end{equation}
which yields

\begin{equation}
T_d^* = \left[ {(d+1) \Gamma(d/2) \over 2 \pi^{d/2}}  {1 \over v^d(\rho) 
\rho p} \right]^{1/(1+d)} \; .
\end{equation}

It has been recently argued \cite{pers} that the presence of a finite 
persistence size
in SW (or a finite persistence time in our case) is a finite size effect
implying that the transition that the system undergoes at $p \to 0$ is first
order. In both SW and DSW, $L^* \sim p^{-\tau}$, with an exponent
$\tau_{\rm SW}=1/d$ in the quenched case, while we have just shown that in 
the dynamical system $\tau_{\rm DSW}=1/(d+1)$. We also expect a first-order 
critical transition for DSW at $p \to 0$. It is not difficult to
understand where the difference in $\tau$ comes from. In DSW, effective jumps,
that is, shortcuts able to carry the activity to an inactive domain, can 
originate in the bulk of the system, and not only at the boundaries of each
active area, as is the case in SW \cite{NW}. The introduction of time
as an additional ``effective'' dimension
naturally leads to the calculated change in the scaling, and establishes
a symmetry between time and space for DSW.

The efficiency of a system with shortcuts can be measured in
terms of the average time that a signal takes to propagate, $T^*$. For
fixed parameters $L$, $p$, and $\rho$, $T^*_{SW} \simeq p^{-1}$, while
$T^*_{DSW} \simeq p^{-1/2}$. Since $p<1$, $T^*_{DSW} < T^*_{SW}$.
In other words, if we are going to design an algorithm to make a signal 
propagate, it would be better to invest effort in dynamical elements able 
to follow arbitrary paths than to construct fixed channels between fixed 
elements.

Dynamic Small-Worlds might be an adequate description for a class of 
systems where the physical movement of the elements is the key ingredient
for the transmission of a signal, and where
a small-world-like behavior may be present. Another example of this class of 
systems may be some species 
of ants, where two different types of individuals, characterized by
two different speeds of movement, are known to coexist \cite{Ants}. This 
moves our model to the field of collective computation, and suggests that a
small-world behavior might be relevant for the overall performance of such 
systems. Some authors have studied epidemics in models where 
elements are able to move on a lattice \cite{Boccara}. It would be interesting
to analyze the dynamics of such a system on a DSW. According to the results 
presented here, they should differ quantitatively from current studies of 
epidemics on SW networks \cite{Dyn}. 
The case $p=0$ of our model closely corresponds to an active process of the
type $A+B \to 2B$ with strong inter-particle interactions. The
spreading of the active front depends on the interplay between the
movement of the active particle at the boundary and the position of
the closest inactive particle in a non-trivial way. This problem is
similar to that of a diffusing prey hunted by $N$ predators \cite{Redner}.

Our model admits a number of generalizations. 
Recent studies have stressed the non-random structure of real networks. 
The scale-free topology of the World Wide Web (WWW), for instance, arises 
from a dynamical
multiplicative process through which a small fraction of nodes accumulates 
most of the links \cite{Bara}.
In the DSW here presented all sites have equal probability of generating or
receiving the next shortcut, resulting in a Poisson distribution of 
shortcuts per site. This rule can be easily modified if our sites are 
provided with memory, such that the 
probability of being selected by an automaton as a landing site depends on
the number of automata which already visited it. This would generate
a skewed distribution of shortcuts, with a small set of frequently visited 
sites. Hubs in the WWW would have their equivalent in `cities' (strongly
prefered sites) in the 
physical space. Actually, the number of potential shortcuts of a real city 
(its population) distributes similarly to the number of connections per node 
in the WWW, and results from a similar growth process \cite{Zan}. The presence
of highly connected nodes in the WWW directly affects the tolerance of the
network against errors \cite{AlCo}. It would be 
interesting to explore if the modification of the landing rule has also further
consequences in the performance of DSW, for instance in the detection of 
the minimal path between two points by an automaton furnished with
local information \cite{Klein}. Further analysis of the DSW model here 
introduced will be the subject of forthcoming works.

\acknowledgments
The authors acknowledge interesting discussions with U. Bastolla,
P. Grassberger, J. Shillcock, N. Shnerb, and D.H. Zanette.
JD was supported by the UPC grant PR98-11.

\end{document}